# Financial density forecasts: A comprehensive comparison of risk-neutral and historical schemes

Ricardo Crisóstomo[a] and Lorena Couso[b]

12 December 2017

## Abstract

We investigate the forecasting ability of the most commonly used benchmarks in financial economics. We approach the usual caveats of probabilistic forecasts studies –small samples, limited models and non-holistic validations– by performing a comprehensive comparison of 15 predictive schemes during a time period of over 21 years. All densities are evaluated in terms of their statistical consistency, local accuracy and forecasting errors. Using a new composite indicator, the Integrated Forecast Score (IFS), we show that risk-neutral densities outperform historical-based predictions in terms of information content. We find that the Variance Gamma model generates the highest out-of-sample likelihood of observed prices and the lowest predictive errors, whereas the ARCH-based GJR-FHS delivers the most consistent forecasts across the entire density range. In contrast, lognormal densities, the Heston model or the Breeden-Litzenberger formula yield biased predictions and are rejected in statistical tests.

[a] Corresponding autor. Comisión Nacional del Mercado de Valores (CNMV), Edison 4, 28006 Madrid, Spain; and National Distance Education University (UNED), Bravo Murillo 38, Madrid, Spain.
*E-mail address*: rcayala@cnmv.es
[b] CaixaBank Asset Management, Madrid, Spain.

## 1. Introduction

Forecasting future asset prices is arguably one of the most relevant problems for risk managers, central bankers and investors. Historical and risk-neutral methods are the most widely used techniques in financial forecasting. Yet, when it comes to evaluate predictions across the entire density range, comprehensive comparisons are scarce and there is no consensus on which models provide better forecasts.

Historical methods generate future predictions based on past prices. These models are easy to implement and extensively used in risk management and stress testing. However, it is well-known that historical patterns do not repeat themselves, particularly in times of economic turmoil. Furthermore, historical models may yield different estimates depending on the length of the calibration window, introducing uncertainty and possible cherry-picking concerns.

Risk-neutral estimates, on the other hand, contain forward-looking expectations and react immediately to changing market conditions, thus being conceptually better suited for forecasting purposes. However, risk-neutral models do not explicitly consider the investors' risk preferences across different future states. Consequently, some agents rapidly dismiss risk-neutral models as the basis for financial predictions.

The previous literature on financial forecasts has been mainly devoted to volatility predictions. Poon and Granger (2003) compare the results from 18 academic papers showing that in 17 of them implied volatilities produce better forecasts than GARCH-based volatilities. Similarly, an extensive survey by Christoffersen, Jacobs, and Chang (2013) find that option-based volatilities beat historical forecasts in most empirical comparisons[1]. Conversely, Canina and Figlewski (1993) find that implied volatilities do not accurately predict the future, providing an exception to the mainstream literature.

Much fewer studies consider entire density forecasts. While empirical analyses tend to find that risk-neutral densities (RNDs) outperform historical-based estimates[2], generalizations to other markets or time-periods are typically limited by three methodological reasons. First, data availability issues have led most researchers to work with relatively small option samples (e.g.: Anagnou et al., 2005 and Liu et al., 2007). Limited samples can significantly impact the evaluation of predictive densities, as the inability to reject a particular model can be due to the low statistical power of the testing procedures (Anagnou et al., 2003).

Second, comparing density estimates from a wide range of schemes requires working with markedly different models and mathematical routines. As a result, most studies have contributed through vis-à-vis comparisons across particular model choices (e.g.: Silva and Kahl 1993; Melick and Thomas, 1997; Alonso, Blanco, and Rubio, 2005) or by surveying specific asset dynamics (Yun, 2014). However, empirical analyses covering a comprehensive range of risk-neutral and historical densities are scarce.

---

[1] Some examples supporting the use of implied volatilities are: Busch, Christensen, and Nielsen (2011), Taylor, Yadav, and Zhang (2010); Giot and Laurent (2007), Jiang and Tian (2005) and Blair, Poon, and Taylor (2001).
[2] See Christoffersen, Jacobs, and Chang (2013).

Third, the validation of financial density forecasts is typically performed through the so-called probability integral transforms (PIT), which assess the statistical consistency between the *ex-ante* densities and the *ex-post* realizations. However, several papers have shown that PIT-based analyses do not consider the forecasting accuracy of the competing methods or the magnitude of its errors, advocating for targeted scoring rules to supplement the PIT assessments[3].

We approach these methodological caveats –small samples, limited models and non-holistic validations– by performing a comprehensive analysis of 15 forecasting schemes during a period of over 21 years. Historical densities are generated using a wide range of models, spanning from returns bootstrapping or standard GARCH dynamics to asymmetric models with filtered historical simulation. Similarly, we estimate RNDs using the most common benchmarks in financial economics, including lognormal densities, stochastic volatility, jump processes and non-parametric distributions.

All density forecasts are evaluated through a 3-tiered criterion. First, we consider a multi-factor goodness-of-fit analysis, assessing each PIT sequence by means of the Berkowitz, Kolmogorov-Smirnov and Jarque-Bera distributional tests. Second, we employ the logarithmic scoring rule, which evaluates the accuracy of each method in predicting the *ex-post* realizations. Third, we are the first to apply, to our knowledge, a return-based Continuous Ranked Probability Score (CRPS) to financial forecasts. The CRPS compares the realizations to the entire *ex-ante* densities, ranking all methods in terms of their prediction errors. Finally, we develop a new indicator, the Integrated Forecast Score (IFS), which aggregates the results from the statistical consistency, local accuracy and forecasting errors analyses into a single composite measure.

We calibrate our RNDs using market-derived option prices only. This approach contrasts with the use of exchange-reported settlement prices, which in many cases are theoretically estimated and already reflect specific modelling choices. Finally, we do not consider in this paper combinations of risk-neutral and historical methods; while this approach seems promising[4], our aim is to shed light on the predictive ability of the most commonly used models in financial economics, thus leaving mixed densities for future research.

The rest of the paper is organized as follows. Section 2 presents the competing models. Sections 3 and 4 contain the dataset and the calibration procedures. Section 5 explains the validation methods, followed by the empirical results in Section 6. Finally, Section 7 concludes**.**

---

[3] See Bao, Lee, and Saltoğlu (2007), Amisano and Giacomini (2007) and Gneiting and Raftery (2007).
[4] See Shackleton, Taylor, and Yu (2010), Høg and Tsiaras (2011), de Vincent-Humphreys and Noss (2012) and Ivanova and Puigvert Gutiérrez (2014).

## 2. Forecasting methods

### 2.1 Historical densities

We employ five specifications for the historical densities. The first assumes that future prices follow a geometric Brownian motion. The corresponding densities are then lognormal. Our second specification generates future price paths by a bootstrapping of past returns, thus randomly drawing returns from the empirical distribution function. For each observation date $t$, the one-day-ahead return is given by:

$$r_{t+1} = \mu + z_{t+1}, \quad z_{t+1} \sim \{r^h\} \tag{1}$$

where $\{r^h\} = (r_1^h,...,r_t^h)$ denotes the set of historical returns and $\mu$ is the daily average return. Next, we consider two standard GARCH(1,1) models. This choice is supported by Hansen and Lunde (2005) which compare 330 ARCH models, finding no evidence that a GARCH(1,1) underperforms more sophisticated dynamics, with the exception of asymmetric models including a leverage effect. Under a GARCH(1,1) future returns are given by

$$\begin{aligned} r_{t+1} &= \mu + e_{t+1} \\ e_{t+1} &= \sigma_{t+1} z_{t+1}, \quad z_{t+1} \sim f_p(0,1) \\ \sigma_{t+1}^2 &= \omega + \alpha e_t^2 + \beta \sigma_t \end{aligned} \tag{2}$$

We consider two specifications for the standardized residuals $z_t$: (i) a Gaussian distribution (GARCH-N) and (ii) a Student's $t$ (GARCH-t). Finally, we evaluate the filtered historical simulation (FHS) approach introduced in Barone-Adesi, Engle, and Mancini (2008), which combines an asymmetric GJR-GARCH model[5] with empirical innovations. Specifically, future returns in the GJR-FHS(1,1) model are computed as

$$\begin{aligned} r_{t+1} &= \mu + e_{t+1} \\ e_{t+1} &= \sigma_{t+1} z_{t+1}, \quad z_{t+1} \sim f_{np}(0,1) \\ \sigma_{t+1}^2 &= \omega + \alpha e_t^2 + \beta \sigma_t + \gamma I_t e_t^2 \end{aligned} \tag{3}$$

where $I_t$ = 1 when $e_t$ < 0 and 0 otherwise, introducing a leverage effect. The innovations $z_t$ are drawn from the scaled empirical distribution function, which is obtained by dividing each estimated return innovation $\hat{e}_t$ by the estimated conditional volatility $\hat{\sigma}_t$[6], thus reflecting the skewness and kurtosis observed in the historical calibration period.

Simulated price paths for all time-series models are generated by drawing from the assumed distribution, calculating the conditional variance (where applicable), drawing another innovation, updating the conditional variance, and so on up to the forecasting horizon, denoted as $t^*$. Consequently, future prices at the forecasting horizon $F_{t^*}$ are given by:

---

[5] Glosten, Jagannathan, and Runkle (1993)
[6] Consequently, the set of scaled non-parametric innovations $f_{np}(0,1)$ is formed by the scaled returns $\{z^h\} = (\hat{e}_1/\hat{\sigma}_1,...,\hat{e}_t/\hat{\sigma}_t)$.

$$F_{t^*} = F_t \exp(\tau\mu + \sum_{i=1}^{\tau} \sigma_{t+i} z_{t+i}) \quad (4)$$

where $F_t$ is the market-observed future price at date $t$, $z_{t+i}$ denotes the daily innovations for each model, and $\tau$ is the number of business days between $t$ and $t^*$.

## 2.2 Risk-neutral densities

Our simplest RND model is again a lognormal specification, but calibrated to match current option prices. It is well-known, however, that the volatility of financial assets is time-variant and stochastic. Therefore, we next consider the Heston (1993) model, which employs a mean-reverting stochastic volatility process that can generate skewed and leptokurtic distributions. Specifically, the risk-neutral dynamics for the Heston model are given by

$$dF_t = \sqrt{V_t} F_t dW_{t,1} \quad (5)$$

$$dV_t = a(\bar{V} - V_t)dt + \eta\sqrt{V_t} dW_{t,2} \quad (6)$$

where $dW_{t,1}$ and $dW_{t,2}$ are two correlated Wiener processes. The empirical literature suggests that a jump component can help explaining the observed equity returns, particularly in short term horizons[7]. Following Bates (1996), we complement the Heston volatility in (6) with a lognormal price jump, thus obtaining the dynamics:

$$dF_t = \sqrt{V_t} F_t dW_t^{(1)} + J_t F_t dN_t - \lambda \mu_J F_t dt \quad (7)$$

where $N_t$ is a Poisson process with intensity $\lambda$ and $J_t$ are the jumps sizes, which are lognormally distributed with an average size $\mu_J$ and standard deviation $v_J$. Leaving diffusion models, we also evaluate the purely discontinuous Variance Gamma (VG) model (Madan, Carr, and Chang, 1998), which combines frequent small moves with rare big jumps. The VG dynamics are:

$$\begin{aligned} F_{t^*} &= F_t e^{\lambda\tau + H(\tau;\sigma,v,\theta)} \\ \lambda &= \frac{1}{v}\ln(1 - \theta v - \frac{\sigma^2 v}{2}) \\ H(\tau;\sigma,v,\theta) &= \theta G(\tau;v) + \sigma G(\tau;v) W_t \end{aligned} \quad (8)$$

where $G(\tau;v)$ a Gamma distribution and the parameters $\sigma$, $v$ and $\theta$ jointly control the volatility, asymmetry and kurtosis.

RNDs for all stochastic models are obtained through characteristic functions. Heston (1993) and Bakshi and Madan (2000) show that the cumulative density function of $F_{t^*}$ can be directly obtained in terms of the characteristic function of $\ln(F_{t^*})$ as follows:

$$CDF_{t^*}(x) = \frac{1}{2} + \frac{1}{\pi}\int_0^\infty \text{Re}\left[\frac{e^{-iw\ln(x)} \psi_{\ln F_{t^*}}(w)}{iw}\right] dw \quad (9)$$

---

[7] See Jones (2003), Bakshi, Cao, and Chen (1997), Cont and Tankov (2004), among others.

where $\text{Re}[\cdot]$ denotes the real operator[8].

Finally, Breeden and Litzenberger (1978) show that given a continuous of non-arbitrable call prices, it is possible to obtain a unique risk-neutral distribution that replicates exactly such option prices. Specifically, we employ the Malz (2014) implementation, which introduces a simple arbitrage-free approach based on cubic spline interpolations across the observed implied volatilities and a flat extrapolation at the endpoints. For each expiry $t^*$, the interpolated volatility function is used to compute the continuous call pricing function $C(x, t^*)$, and these prices are then numerically differentiated to obtain the CDF for all the strikes $x$ as:

$$CDF_{t^*}(x) \approx 1 + e^{r\tau} \frac{1}{\Delta} \left[ C(x - \frac{\Delta}{2}, t^*) - C(x + \frac{\Delta}{2}, t^*) \right] \qquad (10)$$

where $\Delta$ denotes the step size used in the finite differentiation.

---

[8] See Crisóstomo (2014, 2017) for further details.

## 3. Data

### 3.1 Option prices

Our option dataset is comprised of European-style contracts with underlying the IBEX 35 futures. Option prices are observed during a period of over 21 years, from November 1995 to December 2016. To construct our dataset, we follow the recommendations in Christoffersen, Jacobs, and Chang (2013), thus employing liquid option contracts and minimizing the input modelling assumptions.

Our study focuses on strictly market-derived option prices. This choice deviates from the common practice of using exchange-reported settlement prices, which in some cases are theoretically estimated by the exchange even if there is no real activity in the underlying contracts. For instance, prior to expiration, the settlement prices of IBEX 35 options are computed by MEFF assuming a linear relation in the implied volatility function for OTM and ITM options[9]. Therefore, these settlement prices reflect specific modelling choices and using them in the calibration would entail introducing an exogenously-derived volatility shape into all models.

The option dataset is formed by front-month option contracts. This choice is justified by two reasons. First, near-to-expiry contracts exhibit the highest liquidity and thus availability of strictly market-derived prices. Second, the use of monthly cycles allows us to maximize the number of non-overlapping forecasts while minimizing the input modelling assumptions[10]. Observation dates are set 28 calendar days before each monthly expiry date. For such dates, we record the bid and ask prices of all available call and put options. Since ITM options are less actively traded than OTM options, we build our dataset with OTM and ATM calls, whereas OTM put are converted into equivalent call prices using the Put-Call parity.

We only consider options exhibiting contemporaneous bid and ask quotes, while the consistency of each cross-section is ensured by removing contracts that do not respect non-arbitrage conditions[11]. After filtering, we obtain a dataset of 6659 option prices distributed across 254 monthly cycles. The average number of strikes in the cross-sections is 26, ranging from a minimum of 8 to a maximum of 72. Table 1 summarizes the statistics of the option dataset.

---

[9] The slopes are different for ITM and OTM options and are calculated for each maturity. Prior to October 1996, settlement prices were computed using a constant implied volatility (Alonso, Blanco, and Rubio, 2005). MEFF is the official Spanish Futures and Options Exchange.

[10] IBEX 35 options do not exhibit shorter than monthly expirations. Therefore, using shorter periods would lead to forecasting horizons that lack direct option quotes, requiring strong extrapolation assumptions. On the other hand, using longer expiration cycles would entail both reducing the number of non-overlapping periods and relying on less liquid back-month contracts.

[11] Call and put-derived equivalent contracts whose market price is not a convex and decreasing function of the strike are removed from the dataset.

**Table 1: Summary statistics for the IBEX 35 option dataset**

| Option type | Total number | Average per day | Maximum per day | Minimum per day |
|---|---|---|---|---|
| Calls | 3151 | 12 | 38 | 1 |
| Puts | 3508 | 14 | 46 | 3 |
| Overall | 6659 | 26 | 72 | 8 |

| Moneyness | F/K | No. of options | (%) |
|---|---|---|---|
| Deep OTM put | >1.10 | 1755 | 26.36 |
| OTM put | 1.03-1.10 | 1423 | 21.37 |
| Near the money | 0.97-1-03 | 1496 | 22.47 |
| OTM call | 0.90-0.97 | 1541 | 23.14 |
| Deep OTM call | <0.90 | 444 | 6.67 |

Notes: A minimum of 8 options is required to calibrate the Bates model. Therefore, in seven observation dates we supplemented the cross-sections with at-the-money option contracts whose last traded price was consistent with the contemporaneous bid and ask prices. This resulted in an addition of 9 options (0.1% of the sample)

### 3.2 Futures prices

Daily front-month IBEX 35 futures prices are recorded from 19 November 1990 until 23 December 2016[12]. Contrary to options, the daily settlement prices of front-month IBEX 35 futures are computed as the volume weighted average of market transactions between 17:29 and 17:30. The minimum IBEX 35 future price is 1882, recorded on 6 October 1992, and the maximum is 15945.5, attained on 11 December 2007.

Separately, at each monthly expiration date, the final settlement price for each future is determined by MEFF by averaging the spot IBEX 35 index prices from 16:15 to 16:45, taking one observation per minute. These settlement prices constitute the underlying asset of the IBEX 35 options and futures in our study, and hence are used to assess the predictive ability of all forecasting schemes.

### 3.3 Interest rates and dividends

For observation periods between January 1999 and December 2016, we employ the 1-month Euribor. In earlier dates, since the Euribor was not available, we employ the 1-month Mibor. All interest rates are consistently applied in each forecasting period using the corresponding act/360 day count convention. Furthermore, the use of futures contracts have the advantage of making dividend estimation irrelevant; therefore, dividend-related uncertainties do not affect our density forecasts.

---

[12] Prior to 21 July 1992 IBEX 35 futures were not yet traded. For earlier dates, its return is proxied by the spot IBEX 35 returns, whose back-tested prices are available since 29 December 1989.

## 4. Calibration

All model parameters are calibrated on a strictly *ex-ante* basis, considering only the current or historical information available at each observation date.

### 4.1 Historical densities

Historical models can be calibrated to any sufficiently long period of past prices. While this provides flexibility, different calibration windows generate different input values and thus different density predictions. To cope with this uncertainty, we consider two calibration periods: (i) a shorter 6-month window and (ii) a longer 5-year period.

The bootstrapping method does not require any statistical calibration, as it simply entails randomly selecting returns from the relevant historical period. For the lognormal specification, the average return and standard deviation are obtained from the returns in each historical period. Next, ARCH-based parameters are calibrated through maximum likelihood estimation (MLE). Even if the true innovations are not normal, Bollerslev and Wooldridge (1992) show that assuming a Gaussian density for the residuals provides consistent parameter values. Therefore, for the GARCH-N and GJR-FHS models, the parameters are estimated by maximizing the log-likelihood of observing each historical return sequence, which is given by:

$$\ln L = \sum_{j=1}^{t} \left[ -\ln(\sigma_j^2) - \left( \epsilon_j^2 / \sigma_j^2 \right) \right] \quad (11)$$

For the GARCH-t, we employ a two-step estimation process. First, the MLE values of all parameters except $\hat{d}$ are obtained from (11). Then, following Christoffersen (2012), the MLE of $\hat{d}$ is calculated to match the excess kurtosis $\kappa$ of the residuals as $\hat{d} = 6/\kappa + 4$.[13]

Historical parameters are separately estimated for each observation date, thus performing 254 calibrations per model and observation window. Finally, for all time-series models, the calibrated dynamics are used to generate 100.000 price paths, using equation (4).

### 4.2. Risk-neutral densities

RNDs are obtained from the cross-section of option prices at each observation date. For the lognormal model, at-the-money volatilities are computed by linear interpolation of the two nearest-to-the-money implied volatilities in each cross-section. In the Heston, Bates and Variance Gamma models, parameters are calibrated by minimizing the sum of pricing errors. We choose to work with relative errors, which effectively assign a similar weight to all option contracts, generating more consistent results across different strike regions[14]. Denoting $N_t$ the number of option prices are available at date $t$, we estimate the parameter set $\hat{\Theta}$ that minimizes the sum of relative errors for each stochastic model as

$$SRE_t = \sum_{i=1}^{N_t} \frac{\left| C_i - \hat{C}_i(\Theta) \right|}{C_i} \quad (12)$$

where $C_i$ denote the mid-market price of each option in the cross-section and $\hat{C}_i(\Theta)$ is the model-dependent value obtained with the parameter set $\Theta$. All natural constraints on the admissible values of the model's parameters are included in the calibration[15]. For the Variance Gamma process, we also consider the restriction $v^{-1} > \theta + \sigma^2/2$, which is required to avoid numerical blow-ups in the calibration process (see Itkin, 2010 and Crisóstomo, 2017). Following this procedure, risk-neutral parameters are estimated separately for each observation date, performing 254 calibrations per model.

---

[13] We also tried to calibrate the GARCH parameters and $\hat{d}$ simultaneously using the log-likelihood of the sample returns under a GARCH(1,1) with $t(d)$ residuals. However, this approach did not generate improved forecast results.

[14] Despite its popularity, the use of absolute errors entails overweighting the more expensive ITM options while underweighting the cheaper OTM contracts, thus potentially leading to calibration biases.

Finally, for the Malz (2014) implementation of the Breeden-Litzenberger formula, we employ a step size $\Delta = 0.01 F_t$, which avoids negative probabilities in our predictive densities.

---

[15] See Heston (1993), Bates (1996) and Madan, Carr, and Chang (1998) for a description of the admissible parameter values.

## 5. Density forecasts verification

Evaluating financial densities is not particularly straightforward. While in classical inference experiments can be repeated under similar conditions to assess if they conform to a particular distribution, in financial forecasts only one realization is available to evaluate each density prediction. One way to tackle this issue is working with ensemble forecasts, thus jointly assessing a sequence of predictive densities and the corresponding sequence of realizations.

However, even with ensembles, the limited availability of liquid options prices makes it difficult to generate a high number of non-overlapping forecasts. To cope with potential sample issues, we evaluate the predictive ability of all density schemes during a period of over 21 years, hence analyzing 254 non-overlapping monthly cycles. To our knowledge, this is the highest available in comparable research.

Prior to 2007, most predictive densities were evaluated via PIT-based goodness-of-fit analyses. However, PIT analyses are not informative about the accuracy of each competing method or the magnitude of its errors. Therefore, we supplement the goodness-of-fit analyses with two additional scoring rules: (i) the logarithmic score, which evaluates each model accuracy in predicting the final realizations; and (ii) a return-based CRPS, which ranks all density schemes in terms of their prediction errors.

### 5.1 Goodness-of-fit analyses

Diebold, Gunther, and Tay (1998) show that the statistical consistency between a sequence of probabilistic forecasts and the corresponding realizations can be assessed through PIT analyses. For a given date $t$, the PIT represents the quantile of the *ex-ante* distribution at which the *ex-post* realization is observed. Thus,

$$PIT_t = \int_{-\infty}^{x_t^*} f_t(x)\, dx \qquad (13)$$

Intuitively, in a well-specified model, the observed realizations should be indistinguishable from random draws from the predictive distributions, and therefore the sequence of PIT values should be uniformly distributed in the (0, 1) range. However, statistical tests based on uniform variables are typically not powerful enough for small samples (Mitchell and Hall, 2005). Consequently, Berkowitz (2001) proposes a reformulation of the PIT values into a transformed sequence (T-PIT) that should be formed by i.i.d. $N(0,1)$ variables if the predictive densities are correctly specified[16]. The Berkowitz test is carried out by first computing the T-PIT values as $T\text{-}PIT_t = \Phi^{-1}(PIT_t)$ and next the AR(1) model

$$T\text{-}PIT_t - \mu = \rho(T\text{-}PIT_{t-1} - \mu) + \varepsilon_t \qquad (14)$$

is estimated to jointly assess the mean, variance and serial correlation using the likelihood ratio test LR3 = $-2(L(0,1,0) - L(\hat{\mu}, \hat{\sigma}^2, \hat{\rho}))$, which compares the likelihood of the restricted model, where $\mu = 0$, var($\varepsilon_t$) = 1 and $\rho = 0$, with that of an unrestricted one.

---

[16] This reformulation brings about the more powerful tests associated with Gaussian variables.

However, the LR3 test does not directly address the normality of the T-PIT values. If a T-PIT sequence exhibits $\mu = 0$ and var($\varepsilon_t$) = 1, but it is non-normal in its higher moments, Berkowitz's test will fail to detect such failures (Dowd, 2004). Therefore, we complement the LR3 with the Kolmogorov-Smirnov (KS) and Jarque-Bera (JB) tests. The KS test examines whether the maximum distance between the T-PIT distribution and a $N(0,1)$ variable is statistically significant, whereas the JB test considers the skewness and kurtosis of the T-PIT values, assessing the higher moments not covered in the LR3 test.

**5.2 The Logarithmic score**

The accuracy of different forecasting schemes can be compared through the likelihood of the *ex-post* realizations evaluated with the *ex-ante* densities. Following, Liu et al. (2007), Shackleton, Taylor, and Yu (2010) and Høg and Tsiaras (2011) we compute the log-likelihood of the realizations for each predictive scheme as

$$L = \sum_{t=1}^{N} \log(f_t(x_{t*})) \qquad (15)$$

where $f_t$ denotes the *ex-ante* density computed at observation date $t$ and $x_{t*}$ denotes the *ex-post* realization at time $t*$. For each method, the logarithmic rule assigns a loss score to each realization depending on its *ex-ante* probability of occurrence, and by aggregating these scores over the entire sample, density models can be ranked in terms of their out-of-sample accuracy. Furthermore, when all models are potentially misspecified (as it is the case in financial forecasts), the model with maximum $L$ generates the predictive densities which are nearest to the true generating densities, according to the Kullback-Leibler Information Criterion (KLIC)[17] (Bao, Lee, and Saltoğlu, 2007).

**5.3 The Continuous Ranked Probability Score**

The logarithmic score considers the likelihood of the *ex-post* realizations, but ignores any other probability masses. In contrast, the CRPS considers the entire predictive distribution, measuring the statistical distance between the actual realization and all other probabilistic outcomes (Matheson and Winkler (1976)). As a result, the CRPS gives good scores to densities that assign high probabilities to *ex-ante* values that are close, but are not identical, to the one materializing (Gneiting and Raftery (2007)). Denoting by $CDF^m$ and $CDF^r$ the cumulative distributions of the forecasting model and the realization, the CRPS is given by:

$$CRPS_t = \int_{-\infty}^{\infty} \left( CDF^m(x) - CDF^r(x) \right)^2 dx \qquad (16)$$

where:

$$CDF^r(x) = \begin{cases} 0 & for\ x_{t*} < x \\ 1 & for\ x_{t*} \geq x \end{cases} \qquad (17)$$

---

[17] Kullback and Leibler (1951)

Hersbach (2000) shows that the CRPS has the dimension of the parameter $x$, which enters in the calculus through $dx$, thus facilitating the CRPS interpretation as a generalization of the mean absolute error for the entire density forecast.

However, in our empirical sample, the settlement prices of IBEX 35 futures range from a minimum of 1882 to a maximum of 15945.5. Therefore, a CRPS of e.g. 200 index points may have markedly different interpretations depending on the observation date, hindering the comparability of individual CRPS values across different time periods. Consequently, we slightly modify the CRPS to consider return deviations instead of index points. Finally, to aggregate the CRPS values over the entire sample, we compute the average return-based CRPS for each method as:

$$CRPS^{rb} = \frac{1}{N} \sum_{i=1}^{N} \sqrt{\int_{-\infty}^{\infty} \left( CDF^m(x) - CDF^r(x) \right)^2 dx} \qquad (18)$$

For a given forecasting scheme, a $CRPS^{rb}$ of 0.05 can be interpreted as an average return deviation of 5% between the *ex-post* realizations and all the *ex-ante* probability masses, thus providing a direct way to rank competing forecasts.

# 6. Empirical results

Due to the high number of variants tested, we first summarize the different forecasting schemes and their naming conventions. We first consider two simple historical methods: a lognormal density (LN-HIS) and a bootstrapping of historical return (BTS). We also evaluate two standard GARCH models, either with normal or Student's t innovations (GARCH-N / GARCH-t). Finally, we test an asymmetric GJR-GARCH model with filtered historical simulation (GJR-FHS). The required parameters for all historical models are calibrated using either a 6-month period (6m) or a 5-year history (5y).

For the RNDs, we consider again a simple lognormal model (LN-ATM). We then evaluate the density obtained with a stochastic volatility model (HESTON) and a Jump-diffusion process (BATES). Finally, we consider the purely discontinuous Variance Gamma process (VG) and the Malz (2014) implementation of the Breeden-Litzenberger formula (BL-MALZ).

## 6.1 PIT histograms visual inspection

The consistency between an *ex-ante* density scheme and the observed realizations can be intuitively assessed by a simple inspection of the PIT histograms. In a reliable forecast, the histogram of PIT values should resemble a uniform distribution, with departures from the flat line indicating regions where the frequency of realizations is higher or lower than in the *ex-ante* predictions. Figure 1 presents the PIT histograms for all forecasting schemes.

Three main aspects can be highlighted. First, all historical methods calibrated to 6-month periods significantly understate the probability of large losses. This can be seen in the higher occurrence of severe downside movements (i.e.: left-most histogram bar) compared to the *ex-ante* probabilities (i.e.: horizontal dashed line).

Second, although models calibrated to 5-year periods improve the left-tail fit, they conversely allocate too much probability to significant upside movements. The most prominent bias is observed in the GARCH-t(5y), where almost no realizations are observed in the right tail compared to the expected probabilities. In contrast, the GJR-FHS(5y) exhibits a reasonably flat PIT histogram, suggesting that this model generates consistent results across the entire density range.

Third, despite being calibrated to the same inputs, RND methods produce notably different outcomes depending on the model's dynamics. While the Bates model generates a fairly consistent histogram, the HESTON and BL-MALZ models show substantial biases, both allocating too much weight to the left tail and understating either the center or the right tail of the distribution.

## 6.2 Statistical consistency: Goodness-of fit tests

A well-specified model should simultaneously pass the Berkowitz, JB and KS statistical tests, with rejections indicating a departure from the T-PIT normality in either: (i) the mean, sigma or autocorrelation, (ii) the asymmetry and kurtosis or (iii) the distance between the theoretical and the observed CDF. Tables 2 and 3 summarize the results from the goodness-of-fit tests and the T-PIT descriptive statistics.

**Figure 1: Histogram of PIT realizations**

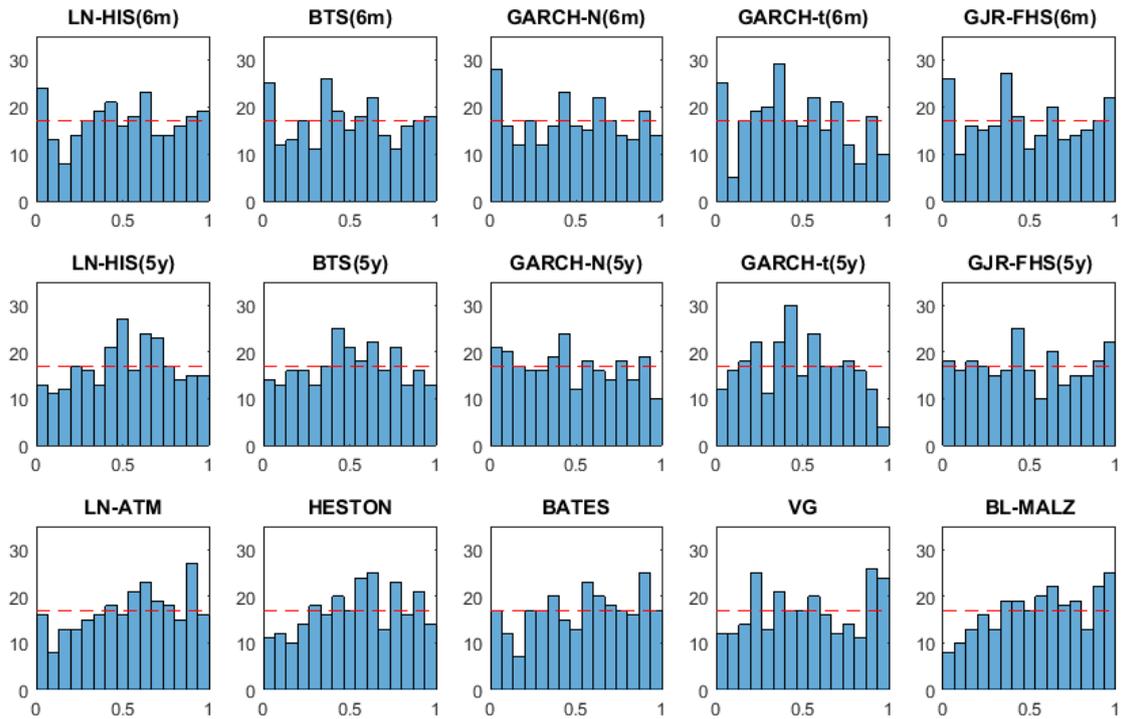

Notes: The horizontal line represents the expected number of realizations in each density region, while vertical bars show the actual number of observations.

### 6.2.1 Historical methods

Of all historical schemes, only the GJR-FHS(5y) and the GARCH-N(5y) simultaneously pass the Berkowitz, JB and KS tests at a 5% significance level. In addition, the GJR-FHS(5y) exhibits more satisfactory *p*-values than the GARCH-N(5y) in all goodness-of-fit tests, thus becoming the most statistically reliable historical model.

The rejection in most historical models (6 out of 10) comes from failures in the JB test, highlighting the key role of assessing the asymmetry and kurtosis of the T-PIT distribution. For the rejected 6-month variants, Table 3 shows that the failure stems from both a negative skewness and a lack of kurtosis. Specifically, where a standard Normal variable should exhibit $\mp$ 1.64 critical values for the $5^{th}$ and $95^{th}$ quantiles, the T-PIT distribution for the rejected variants show critical values of (-1.71, -2.22) and (1.38, 1.78) respectively, exposing (i) a clear underweighting of extreme losses and (ii) a significant asymmetry of results.

Similarly, the LN-HIS(5y) and BTS(5y) failures stem from a left-skewed and leptokurtic T-PIT distribution, whereas the GARCH-t(5y) also performs badly, failing even the Berkowitz test. Although the GARCH-t(5y) rejection can be attributed to the low variance of the T-PIT distribution (0.66), the most striking bias occurs in the right tail, where the fat-tailed Student's t innovations generate an expectation of significant upside movements that is not validated by the *ex-post* realizations. In contrast, all historical methods pass the KS test at a 5% level.

Table 2: Goodness-of-fit analyses and statistical consistency

| Model | Berkowitz | | | | | Jarque-Bera | | Kolmogorov-Smirnov | |
|---|---|---|---|---|---|---|---|---|---|
| | Mu | Variance | Rho | LR3 | $p$-value (%) | Statistic | $p$-value (%) | Statistic (%) | $p$-value (%) |
| **Historical 6-months** | | | | | | | | | |
| LN-HIS(6m) | 0.002 | 1.205 | -0.011 | 4.733 | 19.25 | 7.24 | 3.02 | 5.23 | 47.51 |
| BTS(6m) | -0.045 | 1.217 | -0.012 | 5.770 | 12.33 | 12.54 | 1.79 | 4.10 | 75.07 |
| GARCH-N(6m) | -0.120 | 1.200 | 0.024 | 8.309 | 4.00 | 8.48 | 2.32 | 5.94 | 32.81 |
| GARCH-t (6m) | -0.118 | 0.995 | 0.019 | 3.646 | 30.24 | 6.21 | 4.22 | 6.71 | 19.36 |
| GJR-FHS(6m) | -0.010 | 1.284 | 0.036 | 9.091 | 2.81 | 0.85 | 50.00 | 4.22 | 74.09 |
| **Historical 5-years** | | | | | | | | | |
| LN-HIS(5y) | 0.041 | 0.978 | 0.060 | 1.388 | 70.82 | 57.11 | 0.10 | 8.27 | 5.84 |
| BTS(5y) | 0.001 | 0.948 | 0.065 | 1.375 | 71.14 | 35.95 | 0.10 | 6.50 | 22.41 |
| GARCH-N(5y) | -0.129 | 1.001 | 0.061 | 5.200 | 15.77 | 3.20 | 16.23 | 6.12 | 28.51 |
| GARCH-t (5y) | -0.101 | 0.662 | 0.052 | 21.999 | 0.01 | 4.18 | 9.77 | 8.41 | 5.19 |
| GJR-FHS(5y) | 0.027 | 1.094 | 0.094 | 3.726 | 29.26 | 1.76 | 36.88 | 3.83 | 83.68 |
| **Risk-neutral** | | | | | | | | | |
| LN-ATM | 0.134 | 0.933 | 0.064 | 6.106 | 10.66 | 3.65 | 12.74 | 9.67 | 1.61 |
| HESTON | 0.114 | 0.865 | 0.056 | 6.501 | 8.96 | 43.19 | 0.10 | 10.35 | 0.80 |
| BATES | 0.108 | 1.020 | 0.079 | 4.589 | 20.45 | 1.24 | 50.00 | 7.10 | 14.69 |
| VG | 0.117 | 1.032 | 0.064 | 4.637 | 20.04 | 0.88 | 50.00 | 7.45 | 11.33 |
| BL-MALZ | 0.221 | 0.889 | 0.075 | 15.278 | 0.16 | 0.99 | 50.00 | 10.31 | 0.84 |

### 6.2.2 Risk-neutral methods

Two RND schemes, BATES and VG, simultaneously pass all goodness-of-fit tests, exposing the importance of discontinuous jumps in producing consistent density estimates. In particular, the Bates model exhibits the highest *p*-values in all goodness-of-fit tests, thus being the most statistically reliable RND.

Densities obtained from risk-neutral methods perform better than historical densities in the JB test. As Tables 2 and 3 show, the T-PIT distribution generated by all RNDs, except those of the Heston model, exhibit a skewness and kurtosis that are not statistically different from a standard Normal variable at a 5% level.

In contrast, the LN-ATM, HESTON and BL-MALZ are rejected in the KS test. For these RNDs, the maximum deviation between the empirical and theoretical distributions occurs at CDF points ranging from 0.34 to 0.39, and for KS statistical distances between 9.37% and 10.35%. Therefore, these density schemes allocate too much probability to significant losses compared to the actual realizations. Moreover, the BL-MALZ fails the Berkowitz test (due to its biased 0.22 T-PIT distribution mean), whereas the Heston model is rejected in the JB test (due to negative skewness and a lack of kurtosis in the T-PIT distribution).

**Table 3: Descriptive statistics of the transformed PIT sequences**

| Model | Mean | 5th percentil | Median | 95th percentil | Std | Skewness | Kurtosis | AR(1) |
|---|---|---|---|---|---|---|---|---|
| **Historical 6-months** | | | | | | | | |
| LN-HIS(6m) | 0.002 | -1.972 | 0.011 | 1.779 | 1.100 | -0.302 | 2.396 | -0.011 |
| BTS(6m) | -0.044 | -1.975 | -0.034 | 1.695 | 1.106 | -0.288 | 2.231 | -0.012 |
| GARCH-N(6m) | -0.119 | -2.225 | -0.059 | 1.519 | 1.099 | -0.416 | 2.700 | 0.025 |
| GARCH-t (6m) | -0.133 | -1.708 | -0.154 | 1.328 | 0.965 | -0.265 | 2.251 | 0.038 |
| GJR-FHS(6m) | -0.009 | -1.957 | -0.096 | 1.858 | 1.139 | 0.113 | 2.749 | 0.036 |
| **Historical 5-years** | | | | | | | | |
| LN-HIS(5y) | 0.040 | -1.537 | 0.066 | 1.584 | 0.993 | -0.621 | 0.969 | 0.060 |
| BTS(5y) | -0.001 | -1.686 | 0.033 | 1.492 | 0.980 | -0.554 | 1.365 | 0.065 |
| GARCH-N(5y) | -0.129 | -1.892 | -0.131 | 1.455 | 1.004 | -0.284 | 2.902 | 0.063 |
| GARCH-t (5y) | -0.102 | -1.399 | -0.108 | 1.226 | 0.818 | -0.290 | 2.883 | 0.055 |
| GJR-FHS(5y) | 0.026 | -1.621 | -0.060 | 1.790 | 1.052 | 0.182 | 2.900 | 0.094 |
| **Risk-neutral** | | | | | | | | |
| LN-ATM | 0.134 | -1.649 | 0.202 | 1.613 | 0.970 | -0.294 | 2.913 | 0.063 |
| HESTON | 0.113 | -1.346 | 0.148 | 1.582 | 0.933 | -0.492 | 1.173 | 0.056 |
| BATES | 0.107 | -1.732 | 0.148 | 1.691 | 1.015 | -0.153 | 2.815 | 0.078 |
| VG | 0.116 | -1.455 | 0.065 | 1.806 | 1.020 | 0.139 | 3.059 | 0.064 |
| BL-MALZ | 0.220 | -1.303 | 0.203 | 1.815 | 0.948 | 0.115 | 2.769 | 0.074 |

### 6.3 Local accuracy: Log-likelihood score

Table 4 presents the log-likelihood comparisons for all density schemes. Following Shackleton, Taylor, and Yu (2010) we define the benchmark log-likelihood as the value for the simplest historical method, namely the LN-HIS(6m). For other methods, Table 4 shows the log-likelihood values in excess of the benchmark level. The sample is divided in two subperiods. The first comprises observations dates from November 1995 to December 2006 (134 monthly observations), whereas the second covers from January 2007 to December 2016 (120 observations).

### 6.3.1 Historical methods

The GJR-FHS(5y) and GARCH-N(5y) deliver the highest log-likelihoods among historical methods (+28.74 and +26.29 over the benchmark level). Hence, these models exhibit the greatest accuracy in predicting the *ex-post* realizations and generate the densities which are closer to the true densities according to the KLIC.

Table 4 also reveals two notable findings. First, in all historical schemes the use of 5-year calibration periods results in higher log-likelihoods than employing 6-month windows. Second, for any given calibration length, the log-likelihoods of conditional volatility models are higher than those of lognormal densities or bootstrapped returns, showing that the higher complexity of ARCH models pays off in terms of accuracy.

These results are also robust across subperiods; in all subsamples: (i) the GJR-FHS(5y) achieve the maximum log-likelihood among historical methods and (ii) models calibrated to a 5-year history improve the log-likelihoods of 6-month-based counterparts. In contrast, the period from 2007 to 2016 is challenging for the GARCH-t variants. While the GARCH-t performs well when significant upside or downside movements are observed, this method is a poor predictor of the bulk of *ex-post* realizations, where more modest movements occur. Out of the 59 monthly dates in 2007-2016 where the IBEX 35 returns remain within a ±4% range, in 31 of them either the GARCH-t(5y) or the GARCH-t(6m) assign the smallest probability to the actual realizations. Conversely, in none of such dates these variants exhibit the highest log-likelihood.

**Table 4: Out-of-sample log-likelihoods**

| Model | Observation period | | |
|---|---|---|---|
| | Nov 1995 - Dec 2006 | Jan 2007 - Dec 2016 | Entire Sample |
| **Historical 6-months** | | | |
| LN-HIS(6m) | -971.46 | -872.70 | -1844.16 |
| BTS(6m) | 4.54 | -2.20 | 2.34 |
| GARCH-N(6m) | 12.41 | 0.28 | 12.68 |
| GARCH-t (6m) | 10.23 | -2.59 | 7.63 |
| GJR-FHS(6m) | 10.10 | -0.37 | 9.73 |
| **Historical 5-years** | | | |
| LN-HIS(5y) | 4.21 | 3.08 | 7.29 |
| BTS(5y) | 6.30 | 5.82 | 12.12 |
| GARCH-N(5y) | 17.82 | 8.47 | 26.29 |
| GARCH-t (5y) | 18.39 | 0.99 | 19.37 |
| GJR-FHS(5y) | 19.45 | 9.30 | 28.74 |
| **Risk-neutral** | | | |
| LN-ATM | 15.86 | 5.62 | 21.48 |
| HESTON | 19.87 | 6.88 | 26.75 |
| BATES | 14.62 | 9.41 | 24.03 |
| VG | 19.65 | 11.63 | 31.28 |
| BL-MALZ | 16.20 | 11.49 | 27.69 |

Notes: Log-likelihoods are computed as the value in excess of the LN-HIS(6m) benchmark

### 6.3.2 Risk-neutral methods

Table 4 shows that RND schemes typically exhibit higher log-likelihoods than historical methods. The simplest risk-neutral model, the LN-ATM, achieves a log-likelihood that is higher than in 8 out of 10 historical methods. Furthermore, the VG model delivers the maximum overall log-likelihood (+31.28 over the LN-HIS(6m)), whereas the Bates model also achieves a good score (+24.03).

Despite failing in several goodness-of-fit tests, both HESTON and ML-MALZ deliver notably high log-likelihoods. These results highlight the key role of multi-factor verifications in assessing probabilistic forecasts, exposing how partial evaluations may fail to detect seriously misspecified models. For instance, although BL-MALZ is one of the top-ranked models in local accuracy, we concluded that this method was inconsistent in statistical terms, thus being unable to produce reliable forecasts across the entire density range.

The analysis by subperiods confirms these results and reveals an additional finding. While statistically consistent methods generally show similar log-likelihood gains across different subperiods, inconsistent methods (i.e.: GARCH-t(5Y) or HESTON) exhibit greater variability, indicating that statistical consistency may be associated with a higher log-likelihood stability in different time periods.

**6.4 Forecasting errors: Continuous Ranked Probability Score**

Table 5 summarizes the prediction errors for all forecasting schemes. CRPS values are calculated as the average return-based CRPS for the 21-year sample and the two subperiods, and expressed as the value in excess over the benchmark LN-HIS(6m) level.

RNDs generally exhibit lower forecasting errors than historical densities. Of all 15 density schemes, the VG model achieves the best overall CRPS value (-0.286% compared to the benchmark level) followed by the Bates model (-0.274%). Furthermore, the worst performing RND scheme, the BL-MALZ, outperforms 8 out of 10 historical methods.

The GARCH-N(5y) and GJR-FHS(5y) produce the lowest CRPS among the historical methods (-0.260% and -0.257%, respectively). Moreover, Table 5 confirms the patterns observed in the log-likelihood analyses. First, in all historical schemes the use of 5-year calibration periods results in lower prediction errors. Second, forecasts generated through ARCH-based models typically exhibit lower CRPS values than either lognormal distributions or a bootstrapping of historical returns.

**Table 5: Out-of-sample Continuous Ranked Probability Score**

| Model | Observation period | | |
|---|---|---|---|
| | Nov 1995 - Dec 2006 | Jan 2007 - Dec 2016 | Entire Sample |
| **Historical 6-months** | | | |
| LN-HIS(6m) | 3.672 | 3.801 | 3.737 |
| BTS(6m) | 0.029 | 0.004 | 0.012 |
| GARCH-N(6m) | -0.051 | -0.005 | -0.034 |
| GARCH-t (6m) | -0.090 | 0.065 | -0.021 |
| GJR-FHS(6m) | -0.057 | -0.002 | -0.036 |
| **Historical 5-years** | | | |
| LN-HIS(5y) | -0.210 | -0.209 | -0.214 |
| BTS(5y) | -0.187 | -0.200 | -0.198 |
| GARCH-N(5y) | -0.313 | -0.192 | -0.260 |
| GARCH-t (5y) | -0.268 | -0.079 | -0.184 |
| GJR-FHS(5y) | -0.281 | -0.220 | -0.257 |
| **Risk-neutral** | | | |
| LN-ATM | -0.272 | -0.235 | -0.259 |
| HESTON | -0.282 | -0.184 | -0.241 |
| BATES | -0.293 | -0.242 | -0.274 |
| VG | -0.309 | -0.250 | -0.286 |
| BL-MALZ | -0.210 | -0.215 | -0.217 |

Notes: CRPS figures are computed as the value in excess of the LN-HIS(6m) benchmark and expressed in percentage.

In contrast, the GARCH-t(5y) again provides an exception in 2007-2016. While the GARCH-t is the best performer in all 7 monthly dates where the IBEX 35 futures rise more than 10%, this model produces the highest forecasting errors in 34 of the 42 monthly periods where modest losses or small gains are observed (i.e.: -5% to +1%), making this method the worst-performing among the 5-year historical schemes.

## 6.5 A holistic evaluation: The Integrated Forecast Score (IFS)

To perform a comprehensive evaluation of the competing densities, we develop a novel scoring system that integrates the results from the statistical consistency, local accuracy and forecasting error analyses into a single measure. Following Gneiting and Katzfuss (2014), the IFS assign better scores to predictive methods that exhibit a high forecasting accuracy subject to statistical consistency. To aggregate the individual results, we first normalize the statistical outcomes obtained in the three previous categories into standardized [0, 1] scales.

The normalized score for statistical consistency considers both the number of rejections and the *p*-values obtained in the goodness-of-fit-tests. First, a 0.25 score is allocated to each variant for each non-rejected test. Next, the remainder 0.25 is assigned by averaging the position of each model *p*-values in an empirical [0, 1] scale. Specifically, the best and worst *p*-values in each test are assigned 1 and 0 values, whereas other variants are allocated in the scale through linear interpolation. This scoring rule effectively ranks all forecasting methods in terms of the number of tests passed while further discriminating among competing densities by their *p*-values in the Berkowitz, JB and KS tests.

The normalized scores for local accuracy and forecasting errors assume that the overall log-likelihood and CRPS figures are normally distributed. This can be justified by the central limit theorem: since these values are obtained through a sum of 254 independent and similarly distributed observations, they should converge approximately to a Normal distribution. Specifically, we first obtain the mean and variance of the observed overall values and then each method is ranked in a [0, 1] scale according to their quantile position in the assumed distribution. Finally, the IFS is obtained by averaging, for each density scheme, the three standardized scores.

Table 6 shows the IFS and normalized scores for all forecasting methods. The first aspect to highlight is that RNDs deliver better IFS than historical models. Although we could argue that RNDs stands out due to the low IFS values of the 6-month historical variants, a direct comparison with densities calibrated to 5-year periods confirms the RND's outperformance, with an average IFS of 0.72 versus 0.64 in the 5-year variants. Of all forecasting methods, the VG model achieves the highest IFS (0.880), being the top-rated in local accuracy and forecasting errors, whereas the GJR-FHS(5y) ranks second in the IFS (0.864), and it is the best performer in statistical consistency.

The IFS also validates our previous findings regarding historical methods. First, in all density models the use of 5-year calibration periods results in higher IFS than employing 6-month windows. Second, for any given calibration period, all ARCH-based models achieve better IFS values than either lognormal densities or a bootstrapping of historical returns.

**Table 6: Comprehensive ranking of forecasting schemes**

| Model | IFS | Normalized Scores | | | Calibration |
| --- | --- | --- | --- | --- | --- |
| | | Statistical Consistency | Local Accuracy | Forecasting Errors | |
| **Consitent schemes** | | | | | |
| VG | 0.880 | 0.867 [3] | 0.914 [1] | 0.859 [1] | Risk-neutral |
| GJR-FHS(5y) | 0.864 | 0.929 [1] | 0.868 [2] | 0.793 [5] | 5-year history |
| BATES | 0.817 | 0.871 [2] | 0.747 [6] | 0.833 [2] | Risk-neutral |
| GARCH-N(5y) | 0.812 | 0.823 [4] | 0.811 [5] | 0.802 [3] | 5-year history |
| **Non-consistent schemes** | | | | | |
| LN-ATM | 0.665 | 0.534 [11] | 0.662 [7] | 0.800 [4] | Risk-neutral |
| BL-MALZ | 0.619 | 0,334 [13] | 0.846 [3] | 0.679 [7] | Risk-neutral |
| HESTON | 0.612 | 0.260 [15] | 0.823 [4] | 0.751 [6] | Risk-neutral |
| GARCH-t(5y) | 0.558 | 0.521 [12] | 0.585 [8] | 0.567 [10] | 5-year history |
| BTS(5y) | 0.511 | 0.605 [6] | 0.312 [10] | 0.615 [9] | 5-year history |
| LN-HIS(5y) | 0.476 | 0.588 [8] | 0.170 [13] | 0.670 [8] | 5-year history |
| GJR-FHS(6m) | 0.341 | 0.660 [5] | 0.234 [11] | 0.127 [11] | 6-month history |
| GARCH-t(6m) | 0.280 | 0.561 [10] | 0.178 [12] | 0.103 [13] | 6-month history |
| GARCH-N(6m) | 0.249 | 0.291 [14] | 0.332 [9] | 0.123 [12] | 6-month history |
| BTS(6m) | 0.242 | 0.592 [7] | 0.075 [14] | 0.059 [15] | 6-month history |
| LN-HIS(6m) | 0.232 | 0.574 [9] | 0.048 [15] | 0.072 [14] | 6-month history |

Given the distinct focus of the log-likelihood score and the CRPS (i.e.: local accuracy vs. global errors), it is remarkable that most density schemes achieve either a good or a bad ranking in both. The rationale stems from the relationship between the probability assigned to a given observation and its distance to other probability masses in mound-shaped distributions[18]. However, this relation does not hold linearly in all cases. For instance, while the addition of jumps significantly improves the IFS value of the Bates model compared to Heston[19], its impact in the CRPS and log-likelihood scores goes in opposite directions. This can be attributed to the excessively fat-tailed distributions generated by the Heston model, which assign too much weight to extreme events, thus performing well in the most likelihood-sensitive observations, but it conversely results in higher forecasting errors when the entire *ex-ante* distribution is considered in the computations.

---

[18] In such distributions, realizations falling near the tails (center) of the distributions tend to exhibit (i) a low (high) log-likelihood score and (ii) a high (low) distance to the other probability masses, thus leading to either good or bad ranking in both CRPS and log-likelihood scores.

[19] The IFS gains are driven by the improvement in statistical consistency: through the added jumps, the Bates model simultaneously passes all goodness-of-fit tests, whereas the Heston model fails the JB and KS tests.

## 7. Conclusions

This paper presents a comprehensive analysis of the most commonly used density schemes in financial economics. Through the development of a novel Integrated Forecasting Score (IFS), we show that risk-neutral densities outperform historical-based predictions in terms of information content. The IFS is constructed by aggregating the statistical consistency, local accuracy and forecasting errors results into a single normalized measure.

Using an option dataset covering from 1995 to 2016, we find that the Variance Gamma model simultaneously delivers the largest out-of-sample log-likelihood and the lowest forecasting errors, thus ranking first in the IFS. In contrast, the ARCH-based GJR-FHS achieves the best score in statistical consistency, generating the most reliable forecasts across the entire density range.

We also find two strong patterns regarding historical models. First, in all density schemes the use of 5-year calibration periods outperforms the forecasting ability of 6-month calibration windows. Second, densities obtained from ARCH-type models are more informative than those generated with lognormal methods or a bootstrapping of historical returns. Conversely, frequently used benchmarks like the Heston model or the non-parametric Breeden-Litzenberger formula yield biased predictions and are rejected in statistical tests.

Looking forward, optimally mixing the information content of risk-neutral and historical schemes, and exploring the use of machine learning algorithms to calibrate such models is worthy of research. Moreover, while the IFS provides a simple solution to a complex verification problem, applying the IFS in other datasets or testing its performance in real trading strategies could help to validate the usefulness of this measure as a new tool in financial forecasting. These items remain in our agenda for future research.

## Acknowledgements

We are grateful to MEFF for providing the market data. This research did not receive any specific grant from funding agencies in the public, commercial, or not-for-profit sectors.